\documentclass[a4paper,twoside]{article}

\usepackage{epsfig}
\usepackage{subcaption}
\usepackage{calc}
\usepackage{amssymb}
\usepackage{amstext}
\usepackage{amsmath}
\usepackage{amsthm}
\usepackage{multicol}
\usepackage{pslatex}
\usepackage{apalike}

\usepackage{graphicx}
\usepackage{textcomp}
\usepackage{xcolor}
\usepackage[T1]{fontenc}
\usepackage{cite}

\usepackage{SCITEPRESS}     

\begin{document}

\title{Sociotechnical Challenges of eHealth Technology for Patient Self-Management: A Systematic Review}

\author{\authorname{Stefan Hochwarter\sup{1}\orcidAuthor{0000-0003-2652-135X}}
\affiliation{\sup{1}Department of Computer Science, Norwegian University for Science and Technology, Trondheim, Norway}
\email{stefan.hochwarter@ntnu.no}
}

\keywords{eHealth, mHealth, welfare technology, assistive technology, collaboration, cooperative work, STS, review, challenges}

\abstract{Ageing of society and increase of time spent with chronic conditions challenge the traditional long-term care model. Assistive technology and eHealth are seen to play an important role when addressing these challenges. One prominent example are patient self-management systems. These systems not only transform the way patients with chronic conditions interact with the healthcare system, but also change work practices of care providers. This literature review addresses sociotechnical challenges of eHealth technologies with a strong collaborative component. As a result, four themes are identified and discussed.}

\onecolumn \maketitle \normalsize \setcounter{footnote}{0} \vfill

\section{\uppercase{Introduction}}
\label{sec:introduction}

\noindent Population projections indicate a worldwide population ageing, with high-income countries leading the list~\cite{unWorldPopulationAgeing2017}. This poses a challenge for the delivery of healthcare services, both at a societal and economic dimension. The financial crisis in Europe (European debt crisis) since 2009 and the population ageing forces the countries to rethink their long-term care (LTC) policies~\cite{swartzSearchingBalanceResponsibilities2013}. The rise of life expectancy comes in hand with the increase of disability-adjusted life years (DALYs), whereas non-communicable diseases are leading the list~\cite{oecdHealthGlance20172017}. A shift in primary care for patients with chronic illnesses to address this demographic change is proposed and the use of technology is seen as a powerful mean~\cite{bodenheimerImprovingPrimaryCare2002,bodenheimerImprovingPrimaryCare2002a}.

Welfare Technology (WT), or assistive technology as commonly known outside Scandinavia, is one prominent example of such technology in (primary) healthcare\footnote{In this paper we stick to term Welfare Technology to avoid ambiguity.}. The aim of WT is to increase the life quality and independence of people with physical, psychological or social impairments~\cite{departementenesservicesenterinformasjonsforvaltningInnovasjonOmsorgUtredning2011}. The landscape of WT consists of a wide range of different technologies, such as sensors, Internet of Things (IoT) or GPS. Even though WT often introduces and relies on innovative technical solutions, previous research in the field of WT states that challenges are only 20~\% of technical and 80~\% of organizational nature~\cite{helsedirektoratetVelferdsteknologiFagrapportOm2012}. Hence, the introduction of WT and in turn the digital transformation of healthcare raises challenges of socio-technical nature. This has also an effect on the work of care providers and the communication and collaboration with their patients~\cite{meskoDigitalHealthCultural2017}.

WT involves numerous types of actors which are highly heterogeneous. Platforms are seen as an intermediate between the use of ICT and societal outcomes. They also support communication across the different users of the platform, each with their own, often very specific, requirements to the system. When designing such a platform, the values and needs of these different actors need to be carefully considered and taken into account~\cite{annmajchrzakDesigningDigitalTransformation2016}. The platformization can also lead to a disruption of existing channels for communication between the participating actors of the platform. This can result to uncertainty about where to find information or who is responsible when new information occurs.

The digital transformation also affects the work practices of care providers. In some cases this could also mean the need to perform additional or different work than before the introduction of e.g. WT solutions. New skills are developed by both the care providers and receivers and different forms of communication challenge the traditional form of care~\cite{grisotSupportingPatientSelfCare2018}.

These challenges are common themes within the field of Computer-Supported Cooperative Work (CSCW). The umbrella term CSCW was coined in 1984 during a workshop by Irene Greif and Paul M. Cashman. The terms used to describe this field were not strictly defined, also to  allow broader discussions and invite researchers from various fields to participate. Schmidt and Bannon made an effort to describe those terms in more detail, looking at the meaning of CS (``computer-supported'') and CW (``cooperative work''). They argue that one first needs to understand the underlying mechanisms and nature of cooperative work in order to design computer systems that support cooperative work~\cite{schmidtTakingCSCWSeriously1992}. Two prominent challenges in the field of CSCW are common themes when implementing  welfare technology that changes the ``traditional'' way of working in healthcare, namely the challenge of \textit{disparity in work and benefit} and \textit{disruption of social processes}~\cite{grudinGroupwareSocialDynamics1994}.

The aim of this study to investigate sociotechnical challenges of eHealth technology with a focus on its collaborative nature. Welfare technology poses new challenges to the healthcare system and the way work is organized around the patient, similar to the challenges described in CSCW literature, as stated above. This is especially visible for patient self-monitoring and reporting, as it creates new ways of communication and challenges the traditional division of work of care providers and receivers.

\section{\uppercase{Methods}}
\label{sec:methodology}

\noindent As the eHealth landscape is rapidly changing, and new eHealth solutions and national strategies arise under the umbrella of welfare technology, there is the need for an updated review on the current challenges and opportunities. Even though there have been some systematic literature reviews identified (see for example \cite{vassliAcceptanceHealthRelatedICT2018}), no review was found to understand the collaborative nature of this emerging, and indeed cooperative technology. Hence, concepts were chosen in accordance with the theme, choosing examples of welfare technology that have a high degree of collaboration.

A systematic literature review was conducted in five steps to investigate the study objective to investigate \textit{the sociotechnical challenges of eHealth technology for patient self-management}. The review was designed according to Cruzes and Dyba~\cite{cruzesRecommendedStepsThematic2011}. In a first step a first unstructured search to explore the field and get an understanding of common terms and phrases in this field was conducted. Following, the objective was split up into three concepts, and for each concept similar terms were identifies (see Table \ref{tab:concepts}).

\begin{table}[ht]
  \caption{Mapping the objective to concepts.}\label{tab:concepts}
  \centering
  \small
    \begin{tabular}{lll}
      Concept 1 & Concept 2 & Concept 3 \\
      \hline
      motivation & welfare technology & self-monitoring \\
      pitfalls & ehealth & patient reporting \\
      challenges & mhealth & remote monitoring \\
      & telemedicine & self-management \\
      \hline
    \end{tabular}
\end{table}

\begin{figure}[htbp]
  \centerline{\includegraphics[width=0.15\textwidth]{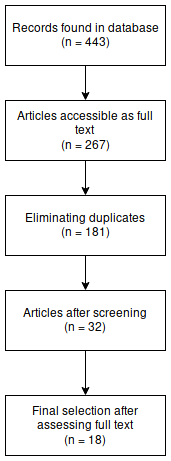}}
  \caption{The study selection process.}
  \label{fig:process}
\end{figure}

Based on these concepts, papers for review were identified searching the Scopus database. Scopus was selected as a primary source as it also includes records from the MEDLINE and EMBASE databases. A complete search history including the number of found and accessible papers can be seen in table \ref{tab:search-history}. Accessible papers were selected based on predefined inclusion and exclusion criteria (Table \ref{tab:inclusion}). In a next steps, duplicates where removed and an initial screening to assess if the paper is relevant to this objective by reading the title and abstract, and evaluating the journal type and research field. After reading the full-text version of the articles identified in the first screening, the final selection of articles for the literature review were selected (see figure \ref{fig:process}).

\begin{table}
  \caption{Inclusion and exclusion criteria.}\label{tab:inclusion}
  \small
  \centering
    \begin{tabular}{l}
      \hline
      original articles published in peer-reviewed journal \\
      articles published in 2009 until 2019 \\
      written in English, German or a Scandinavian language \\
      no study protocols \\
      \hline
    \end{tabular}
\end{table}

Finally, the selected articles were analyzed and concepts were identified and mapped using the reference management system Zotero with the extension ZotFile to extract annotations.

\begin{table*}
  \caption{The search history of the literature review.}\label{tab:search-history}
  \centering
  \tiny
  \begin{tabular}{lll}
      Search & Found & Accessible \\
      \hline
      TITLE-ABS-KEY ( "welfare technology" self-monitoring ) AND DOCTYPE ( ar ) AND PUBYEAR > 2008 & 0 & 0 \\
      TITLE-ABS-KEY ( "welfare technology" patient reporting ) AND DOCTYPE ( ar ) AND PUBYEAR > 2008 & 0 & 0 \\
      TITLE-ABS-KEY ( "welfare technology" remote monitoring ) AND DOCTYPE ( ar ) AND PUBYEAR > 2008 & 1 & 0 \\
      TITLE-ABS-KEY ( "welfare technology" self-management ) AND DOCTYPE ( ar ) AND PUBYEAR > 2008 & 0 & 0 \\
      TITLE-ABS-KEY ( "welfare technology" motivation ) AND DOCTYPE ( ar ) AND PUBYEAR > 2008 & 1 & 1 \\
      TITLE-ABS-KEY ( "welfare technology" pitfalls ) AND DOCTYPE ( ar ) AND PUBYEAR > 2008 & 0 & 0 \\
      TITLE-ABS-KEY ( "welfare technology" challenges ) AND DOCTYPE ( ar ) AND PUBYEAR > 2008 & 6 & 5 \\
      TITLE-ABS-KEY ( ehealth AND self-monitoring AND motivation ) AND DOCTYPE ( ar ) AND PUBYEAR > 2008 & 3 & 2 \\
      TITLE-ABS-KEY ( ehealth AND self-monitoring AND pitfalls ) AND DOCTYPE ( ar ) AND PUBYEAR > 2008 & 0 & 0 \\
      TITLE-ABS-KEY ( ehealth AND self-monitoring AND challenges ) AND DOCTYPE ( ar ) AND PUBYEAR > 2008 & 11 & 8 \\
      TITLE-ABS-KEY ( ehealth AND patient AND reporting AND motivation ) AND DOCTYPE ( ar ) AND PUBYEAR > 2008 & 1 & 1 \\
      TITLE-ABS-KEY ( ehealth AND patient AND reporting AND pitfalls ) AND DOCTYPE ( ar ) AND PUBYEAR > 2008 & 0 & 0 \\
      TITLE-ABS-KEY ( ehealth AND patient AND reporting AND challenges ) AND DOCTYPE ( ar ) AND PUBYEAR > 2008 & 8 & 4 \\
      TITLE-ABS-KEY ( ehealth AND remote AND monitoring AND motivation ) AND DOCTYPE ( ar ) AND PUBYEAR > 2008 & 2 & 0 \\
      TITLE-ABS-KEY ( ehealth AND remote AND monitoring AND pitfalls ) AND DOCTYPE ( ar ) AND PUBYEAR > 2008 & 0 & 0 \\
      TITLE-ABS-KEY ( ehealth AND remote AND monitoring AND challenges ) AND DOCTYPE ( ar ) AND PUBYEAR > 2008 & 17 & 8 \\
      TITLE-ABS-KEY ( ehealth AND self-management AND motivation ) AND DOCTYPE ( ar ) AND PUBYEAR > 2008 & 14 & 8 \\
      TITLE-ABS-KEY ( ehealth AND self-management AND pitfalls ) AND DOCTYPE ( ar ) AND PUBYEAR > 2008 & 0 & 0 \\
      TITLE-ABS-KEY ( ehealth AND self-management AND challenges ) AND DOCTYPE ( ar ) AND PUBYEAR > 2008 & 21 & 12 \\
      TITLE-ABS-KEY ( mhealth AND self-monitoring AND motivation ) AND DOCTYPE ( ar ) AND PUBYEAR > 2008 & 27 & 16 \\
      TITLE-ABS-KEY ( mhealth AND self-monitoring AND pitfalls ) AND DOCTYPE ( ar ) AND PUBYEAR > 2008 & 0 & 0 \\
      TITLE-ABS-KEY ( mhealth AND self-monitoring AND challenges ) AND DOCTYPE ( ar ) AND PUBYEAR > 2008 & 11 & 6 \\
      TITLE-ABS-KEY ( mhealth AND patient AND reporting AND motivation ) AND DOCTYPE ( ar ) AND PUBYEAR > 2008 & 4 & 3 \\
      TITLE-ABS-KEY ( mhealth AND patient AND reporting AND pitfalls ) AND DOCTYPE ( ar ) AND PUBYEAR > 2008 & 0 & 0 \\
      TITLE-ABS-KEY ( mhealth AND patient AND reporting AND challenges ) AND DOCTYPE ( ar ) AND PUBYEAR > 2008 & 13 & 11 \\
      TITLE-ABS-KEY ( mhealth AND remote AND monitoring AND motivation ) AND DOCTYPE ( ar ) AND PUBYEAR > 2008 & 3 & 1 \\
      TITLE-ABS-KEY ( mhealth AND remote AND monitoring AND pitfalls ) AND DOCTYPE ( ar ) AND PUBYEAR > 2008 & 0 & 0 \\
      TITLE-ABS-KEY ( mhealth AND remote AND monitoring AND challenges ) AND DOCTYPE ( ar ) AND PUBYEAR > 2008 & 24 & 15 \\
      TITLE-ABS-KEY ( mhealth AND self-management AND motivation ) AND DOCTYPE ( ar ) AND PUBYEAR > 2008 & 19 & 10 \\
      TITLE-ABS-KEY ( mhealth AND self-management AND pitfalls ) AND DOCTYPE ( ar ) AND PUBYEAR > 2008 & 1 & 1 \\
      TITLE-ABS-KEY ( mhealth AND self-management AND challenges ) AND DOCTYPE ( ar ) AND PUBYEAR > 2008 & 35 & 22 \\
      TITLE-ABS-KEY ( telemedicine AND self-monitoring AND motivation ) AND DOCTYPE ( ar ) AND PUBYEAR > 2008 & 25 & 14 \\
      TITLE-ABS-KEY ( telemedicine AND self-monitoring AND pitfalls ) AND DOCTYPE ( ar ) AND PUBYEAR > 2008 & 1 & 1 \\
      TITLE-ABS-KEY ( telemedicine AND self-monitoring AND challenges ) AND DOCTYPE ( ar ) AND PUBYEAR > 2008 & 28 & 18 \\
      TITLE-ABS-KEY ( telemedicine AND patient AND reporting AND motivation ) AND DOCTYPE ( ar ) AND PUBYEAR > 2008 & 9 & 6 \\
      TITLE-ABS-KEY ( telemedicine AND patient AND reporting AND pitfalls ) AND DOCTYPE ( ar ) AND PUBYEAR > 2008 & 0 & 0 \\
      TITLE-ABS-KEY ( telemedicine AND patient AND reporting AND challenges ) AND DOCTYPE ( ar ) AND PUBYEAR > 2008 & 23 & 13 \\
      TITLE-ABS-KEY ( telemedicine AND remote AND monitoring AND motivation ) AND DOCTYPE ( ar ) AND PUBYEAR > 2008 & 15 & 7 \\
      TITLE-ABS-KEY ( telemedicine AND remote AND monitoring AND pitfalls ) AND DOCTYPE ( ar ) AND PUBYEAR > 2008 & 0 & 0 \\
      TITLE-ABS-KEY ( telemedicine "remote monitoring" challenges ) AND DOCTYPE ( ar ) AND PUBYEAR > 2008 & 31 & 13 \\
      TITLE-ABS-KEY ( telemedicine AND self-management AND motivation ) AND DOCTYPE ( ar ) AND PUBYEAR > 2008 & 49 & 32 \\
      TITLE-ABS-KEY ( telemedicine AND self-management AND pitfalls ) AND DOCTYPE ( ar ) AND PUBYEAR > 2008 & 0 & 0 \\
      TITLE-ABS-KEY ( telemedicine "self-management" challenges ) AND DOCTYPE ( ar ) AND PUBYEAR > 2008 & 40 & 29 \\
      \hline
    \end{tabular}
\end{table*}

\section{\uppercase{Results}}
\label{sec:results}

\noindent Of the initial 443 records found, 18 articles were selected to be included in the literature review. From the 181 accessible articles without duplicates, 159 were excluded mainly because they were not addressing the defined study objective, they were purely focusing on the effect of interventions, or were study protocols for future studies.

\begin{figure}[htbp]
  \centerline{\includegraphics[width=0.35\textwidth]{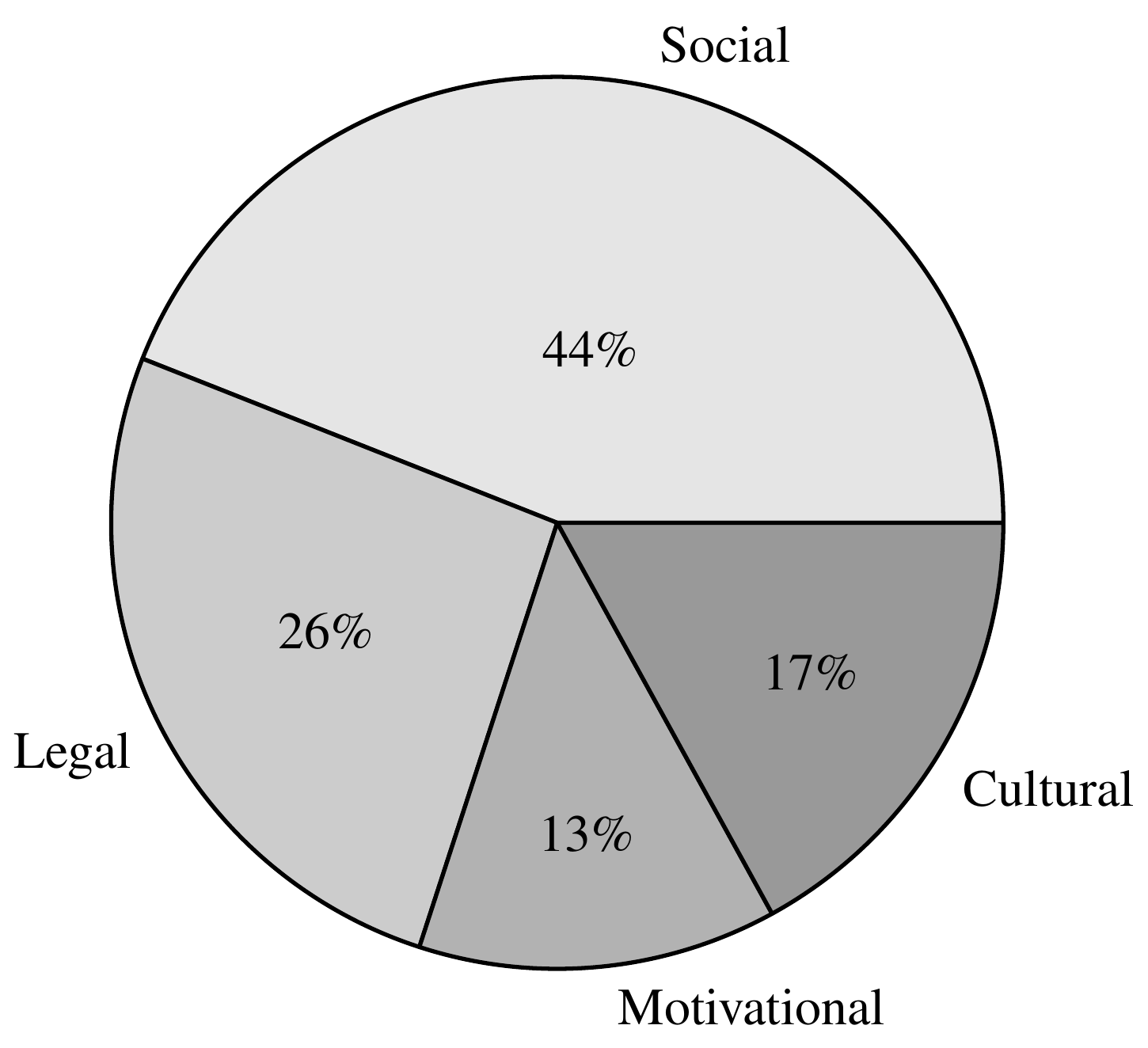}}
  \caption{Identified themes related to study objective.}
  \label{fig:themes}
\end{figure}

Common themes related to the study objective were identified and mapped. The most common theme was related to social pitfalls, followed by motivational and legal challenges. Figure \ref{fig:themes} visualizes the distribution of the themes. Table \ref{tab:results} describes the included papers, their subject of study and the themes identified. 

\begin{table*}
  \caption{Details of reviewed literature and identified themes.}\label{tab:results}
  \centering
  \begin{tabular}{lp{6.3cm}l}
      Reference & Subject of study & Theme(s) \\
      \hline
      \cite{bossuytEhealthInflammatoryBowel2017} & disease management and monitoring for patients with inflammatory bowel disease & legal; social \\
      \cite{dhillonDesigningEvaluatingPatientcentred2016} & patient-centered health management system & cultural \\
      \cite{earlyCaseSeriesOfftheshelf2017} & IT-based self-management of COPD coupled with nurse-coach support & social; motivational \\ 
      \cite{firetMixedFeelingsGeneral2019} & eHealth with pelvic floor muscle training to support self-management of stress urinary incontinence & social \\
      \cite{henkemansDesignEvaluationStartingTogether2018} & pilot implementation of app for self-management & social \\
      \cite{hoaasAdherenceFactorsAffecting2016} & exercise training at home, telemonitoring and self-management, weekly videoconferencing sessions & social \\
      \cite{legido-quigleyChallengesFacingTeleradiology2014} & teleradiology across borders in the European Union & legal \\
      \cite{martinDifferencesReadinessRural2012} & adoption, readiness, and implementation of telemedicine in rural hospitals and primary care providers & legal \\
      \cite{moyanoPerceptionsAcceptabilityText2019} & perception and acceptability of text messaging intervention for diabetes care & social \\
      \cite{nilsenExploringResistanceImplementation2016} & case study of welfare technology describing resistance of implementation & cultural \\
      \cite{odnoletkovaPatientProviderAcceptance2016} & perception of tele-coaching in type 2 diabetes & motivational; legal \\
      \cite{rolloEHealthTechnologiesSupport2016} & eHealth systems for supporting diabetes self-management & cultural; social \\
      \cite{rossDevelopingImplementationStrategy2018} & self-management program for people with type 2 diabetes & motivational \\
      \cite{sanerEHealthCardiovascularMedicine2016} & eHealth and telemedicine challenges and opportunities in cardiology services & legal \\
      \cite{sletteboConflictingRationalesLeader2018} & ethical challenges perceived by leaders
      of community health services for older people & legal; social \\
      \cite{wakeMyDiabetesMyWayEvolvingNational2016} & evaluation of national electronic personal health record and self-management platform for people with diabetes & social \\
      \cite{zibrikPatientCommunityCentered2015} & uptake of eHealth for chronic disease self-management among immigrants and seniors & cultural \\
      \cite{ostlundSTSinspiredDesignMeet2015} & design of welfare technology solutions and addressing the needs of its users - STS-inspired design & social \\
      \hline
    \end{tabular}
\end{table*}

\subsection{Social}

Social challenges and pitfalls are the most common theme. The lack of social support for the patients when using self-management solutions is seen as a major challenge by many articles. Solutions are received as impersonal and complicated by the end-users \cite{rolloEHealthTechnologiesSupport2016,hoaasAdherenceFactorsAffecting2016}. Missing personal contact with the care personnel and hence the absent of physical meetings are reported for long-term care of patients with chronic conditions. Adherence is suffering from this shortcomings~\cite{hoaasAdherenceFactorsAffecting2016,moyanoPerceptionsAcceptabilityText2019}. The lack of personal contact and personal support is the most mentioned challenge~\cite{rolloEHealthTechnologiesSupport2016,hoaasAdherenceFactorsAffecting2016,rolloEHealthTechnologiesSupport2016,firetMixedFeelingsGeneral2019,wakeMyDiabetesMyWayEvolvingNational2016,ostlundSTSinspiredDesignMeet2015}. Stand-alone interventions without personal support are reported to have low usage and acceptance, especially with the older population or minorities~\cite{wakeMyDiabetesMyWayEvolvingNational2016}.

The role of relatives in a healthcare system that is designed around medical conditions rather than the patient's need is another factor for the use of self-management systems, and hence challenges the equitable use and distribution of these systems~\cite{sletteboConflictingRationalesLeader2018}. For children, the role of the parent and their education level is crucial for receiving and effectively using self-management solutions~\cite{henkemansDesignEvaluationStartingTogether2018}.

\subsection{Legal}

Several articles explicitly mention legal barriers for the use of welfare technology. The use of self-management systems challenges the responsibilities of the involved actors. This is reflected by the question of who is responsible in certain situations and how much trust one can put into reports generated by others through self-reporting systems~\cite{legido-quigleyChallengesFacingTeleradiology2014}. Further, the tension between following the minimum legal requirements and the additional care through digital health services challenges the work of care professionals~\cite{sletteboConflictingRationalesLeader2018}. Generally, the lack of a legal framework or the fragmentation of legal frameworks is considered an issue when care providers implement and use welfare technology solutions~\cite{martinDifferencesReadinessRural2012,sanerEHealthCardiovascularMedicine2016,odnoletkovaPatientProviderAcceptance2016}. Finally, legal issues related to privacy, information security and the right to be forgotten hinder the diffusion of WT. These challenges are also linked to social pitfalls~\cite{bossuytEhealthInflammatoryBowel2017,odnoletkovaPatientProviderAcceptance2016}.

\subsection{Cultural}

Cultural differences influence the usefulness of different features of eHealth solutions. Different requirements to the workflow and user interface are reported, based on different cultural background. Features that are of use in one region might not be of use in another region, or culture so to speak~\cite{dhillonDesigningEvaluatingPatientcentred2016}. 
The role of cultural practices, language barriers, and the clash of different professional cultures are stated as inhibited factors. Resistance against the role of co-creators arises out of these factors. Further, poor eHealth literacy is mentioned as a common problem linked to the user's background~\cite{rolloEHealthTechnologiesSupport2016,nilsenExploringResistanceImplementation2016,zibrikPatientCommunityCentered2015}.

\subsection{Motivational}

Finally, pitfalls related to motivational factors were mentioned in three papers. The lack of motivation to use IT rather than poor IT skills is stated as a barrier for patients~\cite{earlyCaseSeriesOfftheshelf2017}. One paper mentions that the motivation of patients with chronic diseases is the biggest issue to overcome when implementing self-management systems.~\cite{odnoletkovaPatientProviderAcceptance2016}.  On the other side, lack of motivation is also reported for care providers who are unwilling to support the system by providing resources~\cite{rossDevelopingImplementationStrategy2018}. 

\section{\uppercase{Discussion}}
\label{sec:conclusion}

\noindent The objective of this paper was to investigate the sociotechnical challenges of eHealth technology for patient self-management. Particularly of interest was the collaborative nature of eHealth technology, hence systems for patient self-management were chosen as a good and timely case to investigate this objective.

The major theme found in this literature review, is the one of social pitfalls. When designing and implementing welfare technology that disrupts existing channels, it has strong social effects. Communication and personal contact between the care providers and patients are subject to change. The shift from care activities from a well-defined and known (at least for the healthcare professionals) environment affects the way care is delivered and experienced. The private homes of the patient play a central role and its perception changes. Also, the way cooperative work is done in the healthcare sector receives an additional component to the already manifold system. Finally, the importance of a peer-network (e.g. relatives, neighbors) is illustrated to ensure personal support. Informal caregivers gain more attention and importance in this setting.

The lack of a legal framework is also mentioned repeatedly as an obstacle to implement WT. In order to avoid additional work without compensation, or work where responsibilities are not clearly defined, a legal framework needs to be established. This also includes GDPR (General Data Protection Regulation) and clear reimbursement systems. Leading the way in the European north, Denmark has established a Digital Health Strategy and an eHealth reference architecture.

Further, it is essential to understand the users and actors involved. The actors can generally not be divided in homogenous groups, rather very heterogenous groups that have distinctive characteristics within the group. Elderly patients often have more than one disorder, and the management of comorbidity is in-turn very individual. This has also been underlined in articles reporting cultural challenges, and calls for a holistic approach that takes this diversity into account when designing WT systems.

Finally, another common theme is the one of motivational challenges. For the care-receivers, in the role of co-creators, motivation can stagnate for chronic conditions where they should self-report even in the absent of symptoms, and this in turn reminds the patients that they are living with a chronic disease. For care providers, on the other hand, poor motivation can be traced back to new work routines, new required skills, and  lack of clearly established boundaries for their responsibilities (cf. legal challenges).

The importance of addressing challenges of collaboration can be seen throughout all four identified themes and their descriptions, although most common within the identified social challenges. The implications of moving care into the home have been addressed by many scholars in the field of CSCW. The lack of personal contact and personal support, as described in section 3.1, can be a consequence of moving care to a different setting. At home, the (invisible) work of relatives for healthcare plays a major role, while their work might not be recognized, or systems and workflows were not designed with them in mind. At the same time, power relations change when healthcare worker visit the patients physically at their homes. So moving care to a new environment challenges the traditional coordination of it~\cite{fitzpatrickReview25Years2013}.

These challenges have been discussed in detail in the field of CSCW, but it aims to go beyond barely describing the context at hand. It is a constructive research domain, it is design oriented and aims to implement better CSCW systems~\cite{schmidtTakingCSCWSeriously1992}. To do so, one must first understand the domain that shall be changed - in our case the cooperative work systems to treat and manage long-term diseases. As Welfare Technology is a rather new umbrella term, which though attracts much attention, the identified challenges are of relevance for the design of new and improved solutions.

\section{\uppercase{Conclusion}}
\label{sec:conclusion}

\noindent This paper focused on the sociotechnical challenges of eHealth technology, and it aims to support the understanding of the complex systems involved. Hence, it can not cover all the aspects that are relevant. The scope of this paper was on the system as a whole, including all actors involved and using self-monitoring or self-management systems as a proxy for systems with high collaborative nature. This is still a very broad perspective, and this makes it easy to miss important details. Future research is recommended to look at a specific group of actors (e.g. patients, policymakers, nurses), at a specific condition (e.g. asthma, dementia) or systems using technologies with unique characteristics (e.g. fall detectors, medicine dispensers). Further, I encourage researchers to become active and uncover what works and what doesn't by using for example action design research or participatory design.

This work has several limitations, mainly due to its scope. First, it uses strict inclusion and exclusion criteria. Only peer-reviewed journal articles are included. To have access to more recent and up-to-date findings, conference papers can further be of interest. Also, only one database was used to retrieve our articles (Scopus). Including other databases which cover a different audience would increase the range of covered topics and perspectives.

Finally, to fully understand this complex and fast-moving field, a different approach, such as a realist review, might bring up more insight into this field.

\vfill

\section*{\uppercase{Acknowledgements}}

\noindent I would like to thank Babak A. Farshchian and Elena Parmiggiani for their constructive feedback. Further, I'd like to thank the three anonymous reviewers for their input that helped to improve this paper.

\bibliographystyle{apalike}
{\small
\bibliography{healthinf2021}}

\end{document}